# Polarity Reversed Robust Carrier Mobility in Monolayer MoS$_2$ Nanoribbons

Yongqing Cai, Gang Zhang,[*] and Yong-Wei Zhang[§]

Institute of High Performance Computing, 1 Fusionopolis Way, Singapore,138632

*Supporting Information Placeholder*

**ABSTRACT:** Using first-principles calculations and deformation potential theory, we investigate the intrinsic carrier mobility (μ) of monolayer MoS$_2$ sheet and nanoribbons. In contrast to the dramatic three orders of magnitude of deterioration of μ in graphene upon forming nanoribbons, the magnitude of μ in armchair MoS$_2$ nanoribbons is comparable to that in monolayer MoS$_2$ sheet, albeit oscillating with width. Surprisingly, a room-temperature transport polarity reversal is observed with μ of hole (h) and electron (e) being 200.52 (h) and 72.16 (e) cm$^2$V$^{-1}$s$^{-1}$ in sheet, and 49.72 (h) and 190.89 (e) cm$^2$V$^{-1}$s$^{-1}$ in 4 nm-wide nanoribbon. The robust magnitudes of μ and polarity reversal are attributable to the different characteristics of edge states inherent in MoS$_2$ nanoribbons. Our study suggests that width-reduction together with edge engineering provide a promising route for improving the transport properties of MoS$_2$ nanostructures.

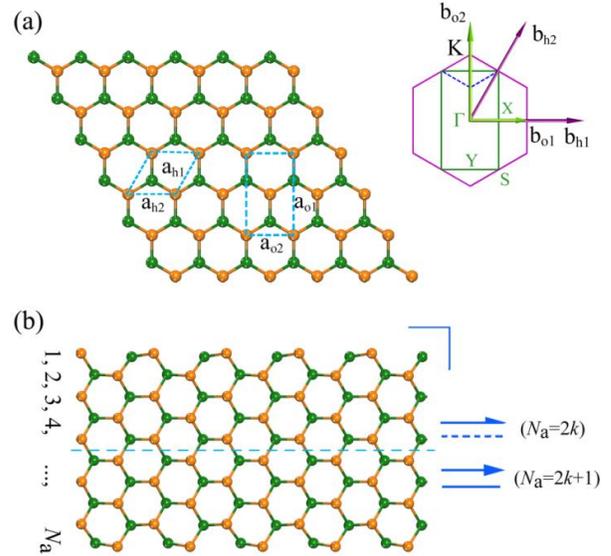

Figure 1. Atomic structure model of monolayer MoS$_2$ sheet and nanoribbons. (a) Atomic structure model of MoS$_2$ monolayer. The dashed lines represent the hexagonal primitive cell (defined by **a**$_{h1}$ and **a**$_{h2}$) and the orthogonal supercell (defined by **a**$_{o1}$ and **a**$_{o2}$). The upper right panel plots the FBZ associated with the two lattices. The blue dashed line shows the folding of the K point in FBZ of the hexagonal lattice into FBZ related to the orthogonal lattice. (b) Atomic structure and nomenclature of armchair MoS$_2$ nanoribbons. The symmetry elements such as the two-fold rotation (screw) axis represented by →, and mirror (glide) planes represented by ⋯, are shown for $N_a=2k+1$ ($N_a=2k$), respectively, where $k$ is an integer. For all the ribbons, there is a mirror plane (represented by ▯) parallel to the ribbon plane.

Two-dimensional (2D) materials are attractive for use in nanoscale electronics and photonics devices owing to their extraordinary unique electronic properties.[1] Graphene is by far the most widely studied 2D material due to its massless charge carriers.[2] However, pristine graphene does not possess a bandgap, a property that is critical for applications in logic transistors to attain a large on/off ratio. Various methods have been proposed to open its bandgap, such as lateral confinement in graphene nanoribbons (GNR).[3] Unfortunately, the carrier mobility ($\mu$) of graphene reduces dramatically upon the bandgap opening. For instance, the magnitude of $\mu$ in a sub-10 nm GNR drops to less than 200 cm$^2$V$^{-1}$s$^{-1}$ (Ref. 4 and 5) from 2×10$^5$ cm$^2$V$^{-1}$s$^{-1}$ in graphene sheet.[6]

More recently, 2D transition metal dichalcogenides (TMDs), composed of atomic layers coupled by van der Waals forces, have gained considerable interest.[7-10] Unlike graphene, single-layer MoS$_2$, a member of TMD family, is a semiconductor with a large bandgap, and hence monolayer MoS$_2$ has been regarded as a promising candidate for field effect transistor (FET) with an on/off ratio exceeding 10$^8$ (Ref. 11 and 12), and high sensitive photodetectors.[13] For a suspended MoS$_2$ sheet, the carrier mobility $\mu$ is found to be in the range of 0.5-3 cm$^2$V$^{-1}$S$^{-1}$ (Ref. 14). By improving sample quality,[15] removing the absorbates[16,17] or depositing atop high-dielectric layer,[12] extrinsic scatters such as charged impurities[18] and grain boundaries[19] can be partially suppressed[20] and the value of $\mu$ can be enhanced to around 200 cm$^2$V$^{-1}$S$^{-1}$.[21] Recently, one-dimensional (1D) MoS$_2$ nanoribbons[22,23] and other TMD materials[24,25] with width down to several nanometers were synthesized (using chemical unzipping of nanotubes, or electron irradiation of sheet). A natural question arises promptly: Is there a similar remarkable reduction (several orders of magnitudes) of $\mu$ in the MoS$_2$ nanoribbons compared to the $MoS_2$ sheet, as occurring in graphene? Also, what is the size-dependent behavior of the mobility? Early studies were focused on the emerging phenomena (for instance, enhanced exciton and photoluminance, indirect-direct bandgap transition) of $MoS_2$ related to the thickness reduction from multi-layers to single-layer.[21] However, effects from the quantum confinement in lateral dimension are still unclear, despite their ultimate importance for nanoelectronics applications.

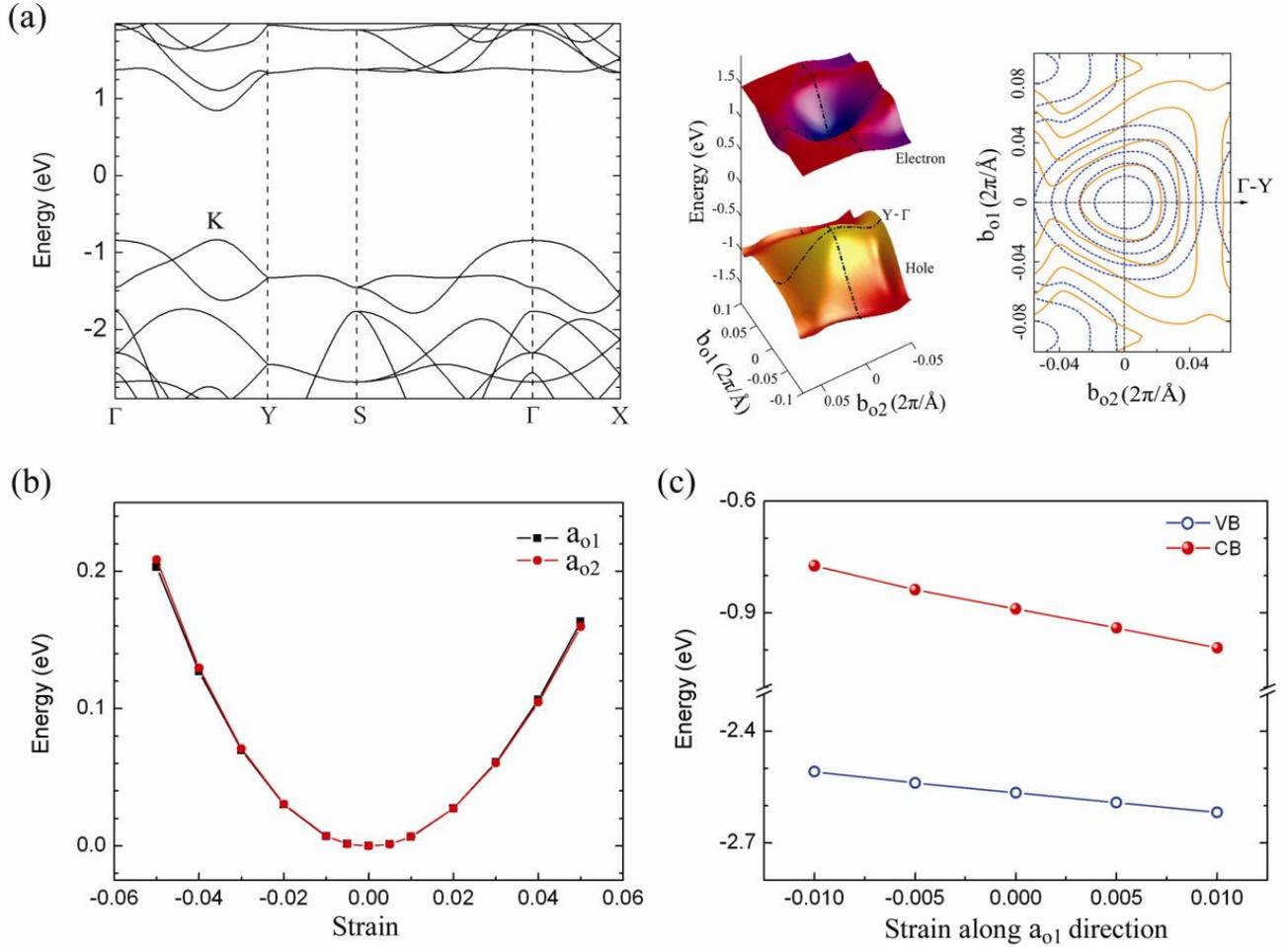

Figure 2. Electronic properties of monolayer $MoS_2$ sheet. (a) Band structure for monolayer $MoS_2$ sheet in the orthogonal supercell (left panel). The surface plots of valleys around VBM and CBM are shown in the middle panel. The dash-dotted lines represent carriers drifting along the $\mathbf{b}_{o1}$ (armchair) and $\mathbf{b}_{o2}$ (zigzag) directions through the K point, which is set to (0, 0). The right panel is the contour plot of the valleys in the 2D $k$-space, which reveals a small anisotropic behavior near the K point for both hole (orange solid lines) and electron (blue dotted lines). The two dotted lines crossing the K point are the in-plane projection of dash-dotted lines in the middle panel. (b) Energy-strain relationship along $\mathbf{a}_{o1}$ and $\mathbf{a}_{o2}$ directions. (c) Shifts of conduction band and valence band under uniaxial strain along $\mathbf{a}_{o1}$ direction.

In this study, by using first-principles calculations and deformation potential theory, we investigate the intrinsic mobility of $MoS_2$ sheet and nanoribbons by means of acoustic phonon scattering mechanism. In stark contrast to graphene, we find that a $MoS_2$ nanoribbon with width down to 4 nm still has a mobility $\mu$ comparable with 2D $MoS_2$ sheet, thus showing another advantage of $MoS_2$ besides the presence of finite bandgap. More interestingly, due to a different scattering intensity between electron and hole, the transport polarity in nanoribbons is reversed from the sheet counterpart: For sheet, electron and hole $\mu$ at room temperature are 72.16 and 200.52 $cm^2V^{-1}s^{-1}$ respectively; while electron and hole $\mu$ are 190.89 and 49.72 $cm^2V^{-1}s^{-1}$ for a ribbon with a width of 4 nm. Analysis of electronic band structure shows that in $MoS_2$ nanoribbons, the non-degraded $\mu$ and reversed polarity are contributed by the edge states. Our study opens up the new possibility of tuning the polarity and/or enhancing the mobility by edge engineering in $MoS_2$ nanostructures.

In inorganic semiconductors, the coherent wave length of thermally activated electrons or holes at room temperature is much larger than the lattice constant, close to that of acoustic phonon modes in the center of the first Brillouin zone (FBZ). The electron-acoustic phonon coupling dominates the scattering of carriers at low energy,[26,27] which can be effectively calculated by the deformation potential (DP) theory proposed by Bardeen and Shockley.[28] Based on the effective mass approximation, the DP theory has been extensively applied to study $\mu$ in 2D [29,30] and 1D [3,31-33] materials with the following forms:

$$\mu_{2D} = \frac{2e\hbar^3 C}{3k_B T |m^*|^2 E_1^2} \quad (1)$$

$$\mu_{1D} = \frac{e\hbar^2 C}{(2\pi k_B T)^{1/2} |m^*|^{3/2} E_1^2} \quad (2)$$

where $m^*$ is the effective mass, $T$ is the temperature. $E_1$ is the DP constant which denotes the shift of the band edges (conduction band minimum for electrons and valence band maximum for holes) induced by the strain. $C$ is the elastic modulus of a uniformly deformed crystal for simulating the lattice distortion activated by the strain. For 2D system, the in-plane stiffness $C^{2D} = [\partial^2 E/\partial \delta^2]/S_0$, where $E$ is the total energy of the supercell, $\delta$ is the applied uniaxial strain, and $S_0$ is the area of the optimized supercell. In the present work, all the above quantities are calculated using density functional theory (DFT) with the generalized gradient approximation (GGA), as implemented in the Vienna Ab initio simulation package (VASP).[34] Spin-restricted calculations are performed using the projector augmented wave method with the Perdew-Burke-Ernzerhof functional (PAW-PBE) and a kinetic energy cutoff of 400 eV. An orthogonal supercell is created for $MoS_2$ sheet as shown in Figure 1a. The atomic plane and its neighboring image are separated by a 20 Å vacuum layer. The $k$-meshes for the sheet and nanoribbons are 10×15 and 1×10, respectively. All the structures are relaxed until the Hellmann-Feynman forces become less than 0.01 eV/Å.

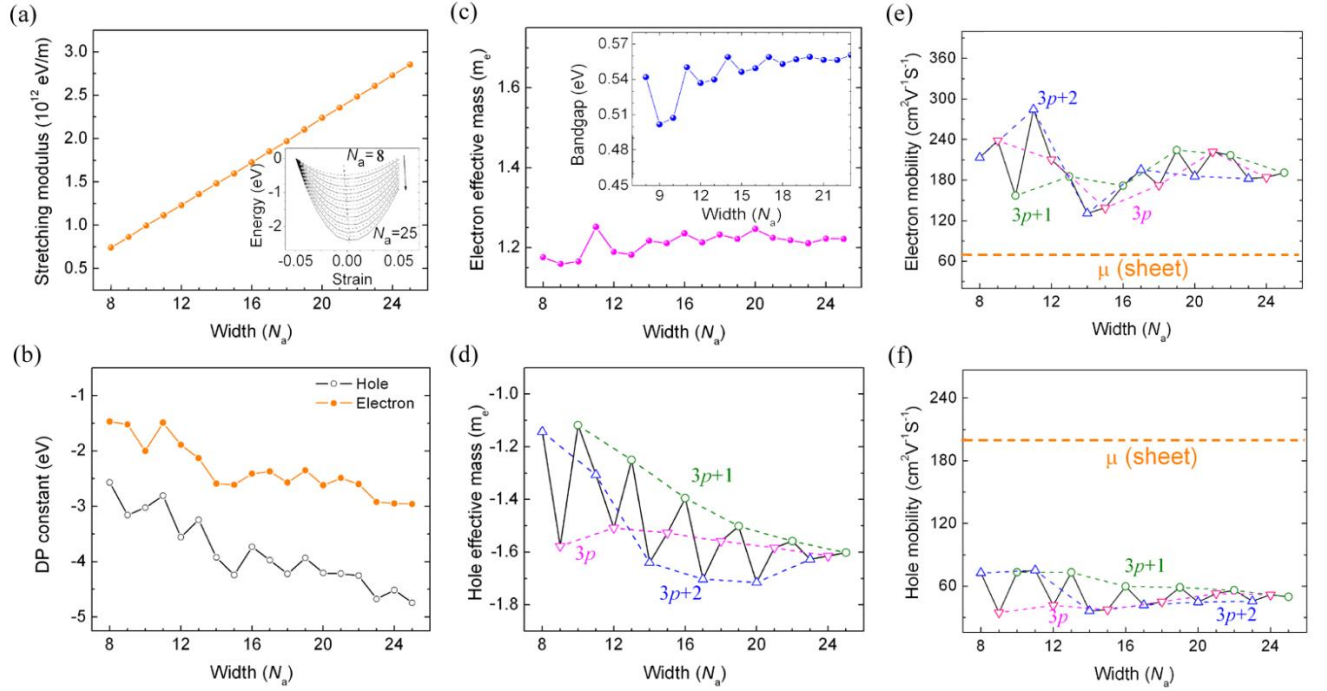

Figure 3. Electronic properties of monolayer $MoS_2$ nanoribbons as a function of width. (a) stretching modulus, (b) deformation potential constant, (c) electron effective mass, (d) hole effective mass, (e) electron mobility, and (f) hole mobility for $MoS_2$ armchair nanoribbons as a function of the ribbon width. The mobility is derived at room temperature (300 K). The inset of (a) shows the relationship between total energy and uniaxial strain along the ribbon direction. The inset of (c) shows the variation of bandgap with ribbon width.

We first investigate the electronic structure and acoustic phonon-limited mobility in monolayer $MoS_2$. The atomic structure is shown in Figure 1a where the orthogonal supercell is enclosed with dashed lines, together with the hexagonal primitive cell for comparison. The supercell built in this way allows for an intuitive demonstration of carrier conduction along the armchair ($\mathbf{a}_{o1}$) and zigzag ($\mathbf{a}_{o2}$) directions. The upper right panel shows the FBZ associated with the hexagonal and orthogonal lattice. Figure 2a shows the band structure for the $MoS_2$ sheet. The K point (fractional reciprocal coordinates: -1/3, 2/3) defined in reciprocal lattice of the primitive cell is folded into (0, 1/3) point sitting at the Γ-Y branch in the FBZ of the supercell (see Figure 1a). The middle panel shows the valleys around the conduction band minimum (CBM) and valence band maximum (VBM), where electrons and holes drifting through K point along the armchair ($\mathbf{b}_{o1}$) and zigzag ($\mathbf{b}_{o2}$) directions are highlighted by the dashed lines. The obtained bandgap of 1.64 eV is consistent with previous DFT studies.[35,36] Through projecting the contour lines into the 2D $k$-space, we observe nearly circular lines near the K center as shown in the right panel, indicating a small anisotropic character in band surface and associated electronic properties of low energy-excited carriers. The effective mass $m^*_{\alpha\beta}$ tensor for charge transport along $\mathbf{a}_{o1}$ and $\mathbf{a}_{o2}$ directions, are calculated by $\hbar^2 [\partial^2 \varepsilon(k)/\partial k_\alpha \partial k_\beta]^{-1}$, which indeed reveals a similar value for holes (0.57 $m_e$ along $\mathbf{a}_{o1}$ direction and 0.60 $m_e$ along $\mathbf{a}_{o2}$ direction) and electrons (0.46 $m_e$ along $\mathbf{a}_{o1}$ direction and 0.48 $m_e$ along $\mathbf{a}_{o2}$ direction). Our results are consistent with previous estimation of 0.64 and 0.48 $m_e$ for holes and electrons, respectively.[37]

**Table 1.** Deformation potential $E_1$, in-plane stiffness $C^{2D}$, effective mass $m^*$, relaxation time $\tau$, and mobility $\mu$ for electron (e) and hole (h) along $\mathbf{a}_{o1}$ and $\mathbf{a}_{o2}$ directions in 2D monolayer $MoS_2$ sheet at 300 K.

| Carrier type | $E_1$(eV) | $C^{2D}$ (N/m) | $m^*$($m_e$) | $\tau$ (fs) | $\mu$ (cm$^2$V$^{-1}$s$^{-1}$) |
|---|---|---|---|---|---|
| e($\mathbf{a}_{o1}$) | −10.88 | 127.44 | 0.46 | 18.89 | 72.16 |
| h($\mathbf{a}_{o1}$) | −5.29 | 127.44 | 0.57 | 64.73 | 200.52 |
| e($\mathbf{a}_{o2}$) | −11.36 | 128.16 | 0.48 | 16.58 | 60.32 |
| h($\mathbf{a}_{o2}$) | −5.77 | 128.16 | 0.60 | 51.83 | 152.18 |

Figure 2b shows the variation of total energy ($E$) with uniaxial strain ($\delta$) applied along $\mathbf{a}_{o1}$ and $\mathbf{a}_{o2}$ directions, respectively. The in-plane stiffness $C^{2D}$ is obtained through fitting the energy-strain curves. The two curves are nearly identical as shown in Figure 2b, and the difference in $C^{2D}$ along the two directions is small (127.44 and 128.16 N/m for $\mathbf{a}_{o1}$ and $\mathbf{a}_{o2}$ directions, respectively). By assuming a finite thickness ($t_0$=0.65 nm) for 2D $MoS_2$ sheet, the three-dimensional (3D) Young's modulus can be estimated as $C^{3D} = C^{2D}/t_0$, amounting to 196.1 GPa and 197.2 GPa, respectively, in good agreement with experimental value (270±100 GPa).[38,39] Figure 2c shows the shift of band edges as a function of strain along $\mathbf{a}_{o1}$ direction and a similar result for $\mathbf{a}_{o2}$ is also observed (not shown). Through dilating the lattice along $\mathbf{a}_{o1}$ and $\mathbf{a}_{o2}$ directions, the DP constant $E_1$ is then calculated as $dE_{edge}/d\delta$, equivalent to the slope of the fitting lines, where $E_{edge}$ is the energy of the conduction (valence) band edge.

Based on the obtained $E_1$, $C^{2D}$, and $m^*$, and using Eq. (1), the acoustic phonon-limited mobility at room temperature (300 K), and the relaxation time ($\tau = \mu m^*/e$) are calculated and compiled in Table 1. The obtained electron mobility is 72.16 ($\mathbf{a}_{o1}$ direction) and 60.32 cm$^2$V$^{-1}$S$^{-1}$ ($\mathbf{a}_{o2}$ direction), respectively. For holes, the mobility is about three times larger than that of electrons, with a value of 200.52 ($\mathbf{a}_{o1}$ direction) and 152.18 cm$^2$V$^{-1}$S$^{-1}$ ($\mathbf{a}_{o2}$ direction), respectively. The calculated mobilities are in good agreement with experimental values.[14-19] In particular, our prediction that the hole mobility is about three-fold larger than electron mobility is consistent with a recent Hall effect measurement, which found that the hole mobility is twice the value of the electron mobility.[40] According to our calculation, the larger hole mobility originates from the smaller deformation potential constant $E_1$ of the valence band compared with that of the conduction band.

We next investigate $\mu$ of $MoS_2$ nanoribbons. It is well-known that the edge states exist in $MoS_2$ nanoflakes and dominate their catalytic performance.[41] However, the role of edge states in the dynamics of carriers is still unclear. As the zigzag $MoS_2$ nanoribbons are metallic,[42,43] here we only consider the semiconducting armchair $MoS_2$ nanoribbons, which are classified by the number of Mo-S dimer lines ($N_a$) across the ribbon width as shown in Figure 1b. A series of armchair nanoribbons with width ranging from 7≤$N_a$≤25 (from 1.1 nm to 4.0 nm) are calculated. For 1D case, the stretching modulus is defined as $C^{1D}=[\partial^2 E/\partial\delta^2]/L_0$, where the uniaxial strain $\delta$ is applied along the ribbon direction, and $L_0$ is the lattice constant of the optimized ribbon. Figure 3a shows that $C^{1D}$ increases continuously with $N_a$. Assuming a finite width ($W_0$) of the optimized ribbon, the effective $C^{2D}$ for these ribbons can be obtained as $C^{1D}/W_0$, ranging between 88.9-114.8 N/m (see Supporting Information).

Figure 3b shows the DP constant $E_1$ due to the quasi-static deformation of the ribbon. It is seen that $|E_1|$ of hole is about two times larger than that of electron. An opposite trend is true for the 2D sheet where electron tends to experience stronger acoustic phonon scattering (Table 1). It should be noted that with further increase in the width of the ribbon, the values of $|E_1|$ for hole and electron are not converged to those of the 2D sheet. This is due to a dramatic difference in the character of frontier orbitals between nanoribbons and sheet, which will be discussed below.

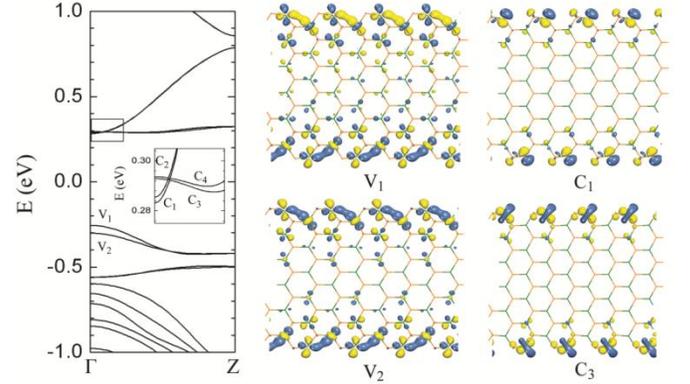

Figure 4. Band structure and frontier orbitals of $MoS_2$ nanoribbons with $N_a$=12. The Fermi level is set to zero. The inset in the left panel is a close-up view of the conduction band. The right panels present the frontier orbitals at the Γ point.

Figure 3c and 3d present the effective masses for electron and hole in the nanoribbons. The electron effective mass increases gradually and then levels off to a value of around 1.2 $m_e$ for $N_a$>15 (see Figure 3c). Interestingly, for hole, the effective mass shows a strong oscillation where three distinct families (3$p$, 3$p$+1, and 3$p$+2) of the ribbons can be clearly identified (see Figure 3d). We also calculate the bandgap (Figure 3c inset) of the $MoS_2$ nanoribbons where the bandgap value also oscillates initially and then converges to a constant value of 0.56 eV. Similar behavior is well-known in the armchair GNR, where a width-dependent oscillation of bandgap occurs.[2] However, in GNR, there is no appreciable width-dependence of effective mass, and the electrons and holes possess almost the same effective mass.[32] The oscillating behavior in the effective mass of $MoS_2$ nanoribbons can be explained by the symmetric characteristics of frontier orbitals which will be discussed shortly. In addition, for wide $MoS_2$ nanoribbons ($N_a$>20), the holes have a much larger effective mass (1.6 $m_e$) than electrons (1.2 $m_e$). This is in contrast to the 2D sheet, which shows a similar effective mass between holes and electrons (Table 1).

**Table 2. Assignment of irreducible representation of $C_{2v}$ group to $V_1$ and $V_2$ frontier orbitals at $\Gamma$ for $MoS_2$ armchair nanoribbons ($N_a$=8-15).**

|       | 8   | 9   | 10  | 11  | 12  | 13  | 14  | 15  |
|-------|-----|-----|-----|-----|-----|-----|-----|-----|
| $V_1$ | $B_1$ | $B_1$ | $A_1$ | $B_1$ | $B_1$ | $A_1$ | $B_1$ | $B_1$ |
| $V_2$ | $A_1$ | $A_1$ | $B_1$ | $A_1$ | $A_1$ | $B_1$ | $A_1$ | $A_1$ |

The room temperature $\mu$ for ribbons is shown in Figure 3 (e) and 3(f). The mobility initially oscillates with the width of the nanoribbons and then levels off to a value (for $N_a$>23) of around 190.89 and 49.72 $cm^2v^{-1}s^{-1}$ for electrons and holes, respectively. For holes, there are three families of the ribbons showing smooth trends with the $3p+1$ family having the largest $\mu$. For electrons, a larger $\mu$ is observed, which arises from the much smaller DP constant and the generally smaller effective mass compared with those of holes. In Figure 3e and 3f, $\mu$ in 2D sheet, represented by the horizontal dashed orange line, is also plotted for comparison. Surprisingly, a reversal of transport polarity is observed in $\mu$ when 2D monolayer sheet is cut into 1D nanoribbon. In monolayer $MoS_2$ sheet, the holes show a three-fold larger mobility than the electrons, while this polarity is reversed in nanoribbons. Our results may account for the depletion-to-enhancement mode transition recently observed in FET based on $MoS_2$ nanoribbons (the channel width from 2 μm down to 60 nm).[44] It was shown that through solely reducing the width of the ribbons, the transport characteristic of the FET changes clearly from depletion mode to enhancement mode. In traditional bulk semiconducting devices, this mode transition can only be achieved by chemical doping. However, stable chemical doping in monolayer channel is generally difficult. Our predicted polarity tunability of carriers in single-layer $MoS_2$ through cutting nanoribbons, together with proper edge engineering, provides a promising avenue to enhance the performance of nanoelectronics devices.

It is well-known that the $\mu$ of graphene sheet is dramatically reduced by around three orders of magnitude upon cutting into nanoribbons.[4,5,32] However, our present work reveals that the mobility of $MoS_2$ nanoribbons is comparable to, or even larger than that of monolayer $MoS_2$ sheet. The width-insensitive carrier mobility is another appealing feature of $MoS_2$ for applications in nanoscale electronics and photonics devices.

To understand the staggering behavior of the effective mass and $\mu$ with width in $MoS_2$ nanoribbons, we analyze the spatial distribution of frontier orbitals. Since all the nanoribbons have a similar electronic character, the $N_a$=12 nanoribbon is chosen as a representative for analysis. Figure 4 shows that the ribbon is a direct bandgap semiconductor. A wealth of edge states are clearly distributed between -0.5 to 1 eV within the bulk bandgap of $MoS_2$. Two valence bands, labeled as $V_1$ and $V_2$, and two conduction bands, labeled as $C_1$ and $C_2$, are identified (inset of Figure 4) where all of them have a parabolic character. In addition, two relatively flat bands ($C_3$ and $C_4$) are located slightly above the $C_1$ and $C_2$ bands. Atomic orbital analysis shows that all these states are mainly composed of Mo 4$d$ manifolds: $V_1$ with in-plane $d_{x^2-y^2}$ and $d_{z^2}$ orbitals; $C_1$ with out-of-plane $d_{xy}$ orbital; and $C_3$ with $d_{z^2}$ orbital. The $V_2$, $C_2$ and $C_4$ states are geometrically same with $V_1$, $C_1$ and $C_3$, respectively, except a phase difference.

It is interesting to compare the width-dependent behavior between GNR and $MoS_2$ nanoribbons. For the GNR, the frontier orbitals are well extending over the whole ribbon and the $\mu$ shows a variation of 2~3 orders of magnitude with width.[32] In contrast, for $MoS_2$ nanoribbons, despite evident edge coupling for narrow ribbons, all the states are mainly localized at edges, thus the $\mu$ tends to be less sensitive to width. The carriers at edges of sub-10 nm ribbons show a comparable $\mu$ with that in $MoS_2$ sheet.

According to group theory, the electronic eigenstates at $\Gamma$ point can be classified as the irreducible representation (IR) of the corresponding little group. The $MoS_2$ nanoribbons have a $C_{2v}$ symmetry (Figure 1). In Table 2, we list the symmetrical characteristics for $V_1$ and $V_2$ states at $\Gamma$. Both states can be assigned to an $A_1$ ($B_1$) IR showing an even (odd) characteristic of the orbital under mirror or glide operation. Interestingly, there is an alternating sequence of the IR character of the states with respect to width of the ribbon. The $3p+1$ family, which has the smallest hole effective mass and the largest hole mobility, shows an $A_1$ character. The $3p$ and $3p+2$ families have a $B_1$ character. The staggering oscillatory behavior of hole effective mass and mobility can thus be attributed to the different symmetric characteristics of the edge states, which result in different carrier-phonon coupling. As expected, the conduction bands (virtual orbitals) are generally more difficult to be described theoretically than the valence bands (occupied orbitals).[45] Although the DP constant derived from the shifts of CBM and VBM tends to be less affected by this issue, the curvature of the CBM may not be accurately described by the DFT used here. Therefore, there is no clear ordering behavior in effective mass and mobility for electrons.

Finally, we would like to demonstrate that the spatial distribution and orbital composition of the edge states in the $MoS_2$ nanoribbons dominate the carrier mobility. For nanoribbon with $N_a$=12, the $V_1$ band is a $d_{x^2-y^2}$ and $d_{z^2}$ hybridized state extending along the edge direction, whereas the $C_1$ level with $d_{xy}$ orbital plane is more localized and perpendicular to the ribbon plane. Therefore, the $V_1$ level tends to shift more under elastic deformation along the ribbon direction according to Eq. (2), thus leading to a larger DP constant and a smaller mobility for hole compared with those for electron populating at $C_1$ level. This indicates that the carrier mobility of $MoS_2$ nanoribbons can be modulated by tuning the distribution of edge states through edge engineering (for example chemical functionalization). For GNR, the edges are stabilized by passivation of hydrogen atoms due to strong C-H bond, thus largely impeding adsorption of other molecules. However, for $MoS_2$ nanoribbons, H-passivated edges are less stable as the H atoms tend to be desorbed from the edges due to the relatively weaker S-H and Mo-H bonds.[46] The H-free edges in $MoS_2$ allow more flexibility in functionalizing the edges through anchoring external atoms or molecules. This can in turn affect the distribution of edge states and may decrease the scattering, thus promoting the mobility.

In conclusion, we have studied the acoustic phonon-limited carrier mobility for both monolayer $MoS_2$ sheet and nanoribbons. In contrast to graphene where a remarkable reduction of carrier mobility occurs in nanoribbons, we find that there is no marked reduction of carrier mobility in $MoS_2$ nanoribbons compared to $MoS_2$ sheet. This character renders $MoS_2$ an ideal material for use in nanoelectronics without losing the mobility upon width reduction. Moreover, we find that the spatial distribution and orbital composition of the edge states dominate the dilation behavior of conducting bands. As the edge states are less sensitive to the width variation and persist in the gap of $MoS_2$ sheet, our study suggests that edge engineering by chemical functionalization or strain in $MoS_2$ nanostructures can be an effective avenue to tune the transport property. It is expected that the robustness of mobility in ultra-narrow $MoS_2$ nanoribbons can lead to a wide range of applications in nanoelectronics devices.

## ASSOCIATED CONTENT

### Supporting Information

Irreducible representation of $C_{2v}$ group, stretching modulus of $MoS_2$ nanoribbons, band structure obtained by HSE06 calculation, strain effect on the effective mass of $MoS_2$ nanoribbons. This material is available free of charge via the Internet at http://pubs.acs.org.

## AUTHOR INFORMATION

### Corresponding Author

*zhangg@ihpc.a-star.edu.sg;  §zhangyw@ihpc.a-star.edu.sg

### Notes

The authors declare no competing financial interests.

## ACKNOWLEDGMENT

The authors gratefully acknowledge the financial support from the Agency for Science, Technology and Research (A*STAR), Singapore and the use of computing resources at the A*STAR Computational Resource Centre, Singapore.

TOC

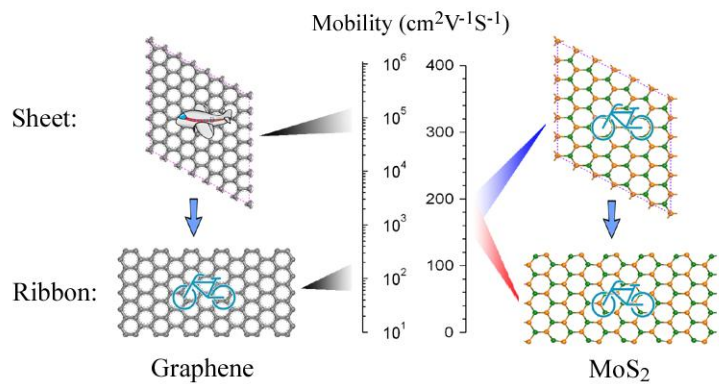